\documentclass[sigconf]{acmart}

\usepackage{subcaption}
\usepackage{caption}
\usepackage{booktabs}
\usepackage{tabularx}
\usepackage{graphicx}
\usepackage{xcolor}

\AtBeginDocument{%
}

\setcopyright{acmlicensed}
\copyrightyear{2026}
\acmYear{2026}
\setcopyright{cc}
\setcctype{by}
\acmConference[CHI EA '26]{Extended Abstracts of the 2026 CHI Conference on Human Factors in Computing Systems}{April 13--17, 2026}{Barcelona, Spain}
\acmBooktitle{Extended Abstracts of the 2026 CHI Conference on Human Factors in Computing Systems (CHI EA '26), April 13--17, 2026, Barcelona, Spain}
\acmDOI{10.1145/3772363.3799099}
\acmISBN{979-8-4007-2281-3/2026/04}

\begin{document}

\title{The Topology of Recovery: Using Persistent Homology to Map Individual Mental Health Journeys in Online Communities}

\author{Joydeep Chandra}
\affiliation{%
  \institution{BNRIST, Dept. of CST, Tsinghua University}
  \city{Beijing}
  \country{China}
}
\email{joydeepc2002@gmail.com}

\author{Satyam Kumar Navneet}
\affiliation{%
  \institution{Independent Researcher}
  \city{Bihar}
  \country{India}
}
\email{navneetsatyamkumar@gmail.com}

\author{Yong Zhang}
\affiliation{%
  \institution{BNRIST, Dept. of CST, Tsinghua University}
  \city{Beijing}
  \country{China}
}
\email{zhangyong05@tsinghua.edu.cn}

\begin{teaserfigure}
  \includegraphics[width=\textwidth, trim=0 0 0 1.5cm, clip]{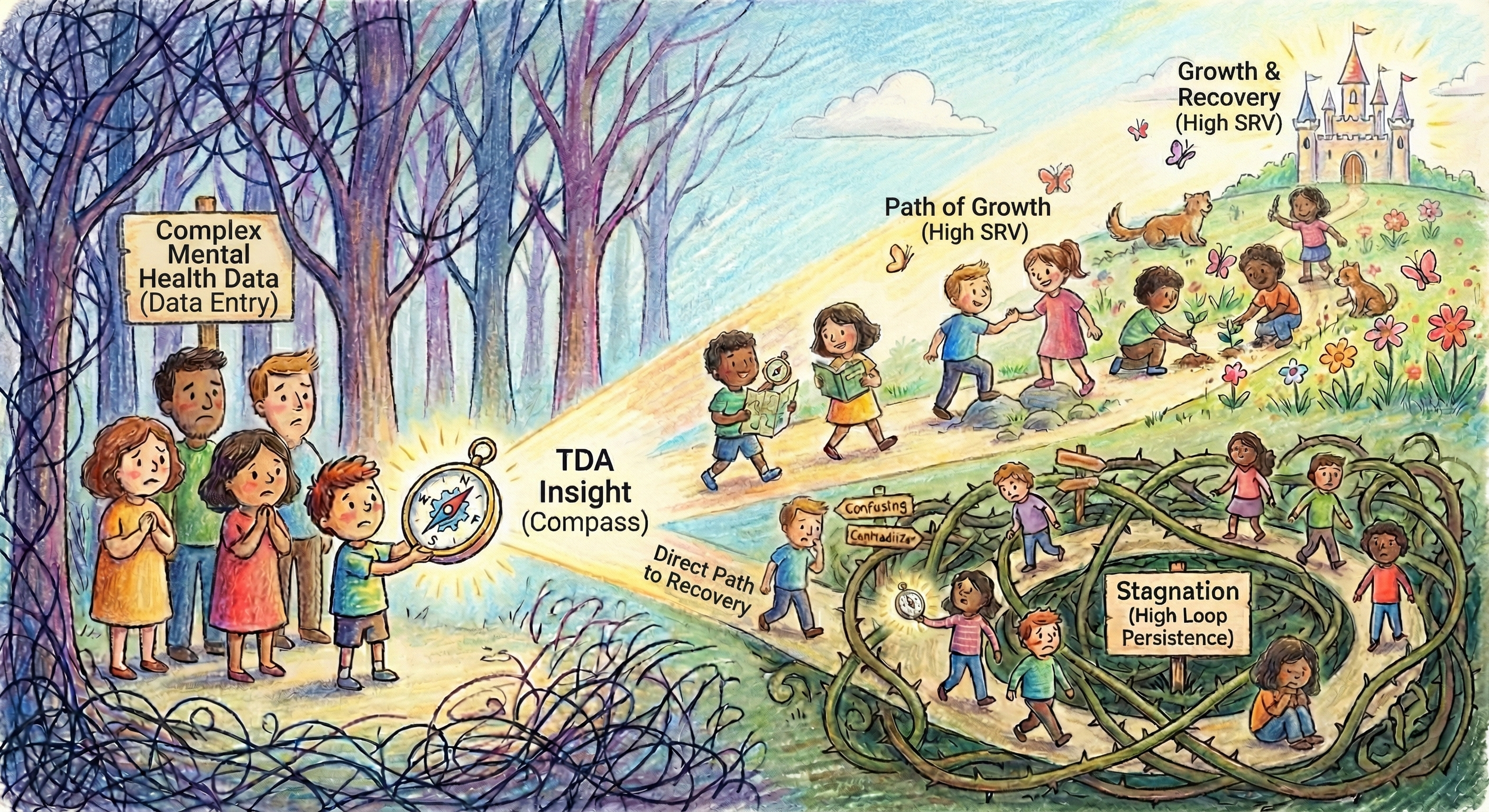}
  \caption{A conceptual illustration of our Topological Data Analysis (TDA) framework. Our approach acts as a computational "compass" to navigate complex longitudinal mental health data. By mapping individual journeys as trajectories through semantic space, we can distinctively map users experiencing "stagnation" (characterized by topological loops \& returning to distress) versus those on a path of "growth \& recovery" (characterized by semantic flares \& a high Semantic Recovery Velocity).}
  \Description{Illustrated landscape scene depicting the Topological Data Analysis framework as a metaphorical journey. On the left, a group of cartoon figures enters through an area labeled 'Complex Mental Health Data (Data Entry),' representing raw longitudinal posting data. In the center, a large compass labeled 'TDA Insight (Compass)' serves as the analytical tool, with a path labeled 'Direct Path to Recovery' extending from it. A signpost nearby reads 'Confusing Semantics.' The landscape diverges into two distinct paths: one leading to the upper right where cheerful figures celebrate at a hilltop labeled 'Growth \& Recovery (High SRV),' reached via an upward trail labeled 'Path of Growth (High SRV),' representing users whose semantic trajectories flare outward into new topics over time. The other path curves downward to the lower right where figures walk in circles in a valley labeled 'Stagnation (High Loop Persistence),' representing users whose posting trajectories form topological loops, repeatedly returning to similar distress-focused semantic regions. The scene uses a watercolor-style landscape with trees, clouds, and winding paths to convey that mental health recovery is a complex, non-linear journey that TDA can help navigate and distinguish between stagnation and growth patterns.}
  \label{fig:teaser}
\end{teaserfigure}

\renewcommand{\shortauthors}{Joydeep Chandra et al.}

\begin{abstract}
Understanding how individuals navigate mental health challenges over time is critical yet methodologically challenging. Traditional approaches analyze community-level snapshots, failing to capture dynamic individual recovery trajectories. We introduce a novel framework applying Topological Data Analysis (TDA)—specifically persistent homology to model users' longitudinal posting histories as trajectories in semantic embedding space. Our approach reveals topological signatures of trajectory patterns: loops indicate cycling back to similar states (stagnation), while flares suggest exploring new coping strategies (growth). We propose Semantic Recovery Velocity (SRV), a novel metric quantifying the rate users move away from initial distress-focused posts in embedding space. Analyzing 15,847 r/depression trajectories \& validating against multiple proxies, we demonstrate topological features predict self-reported improvement with 78.3\% accuracy, outperforming sentiment baselines. This work contributes: (1) a TDA methodology for HCI mental health research, (2) interpretable topological signatures, and (3) design implications for adaptive mental health platforms with ethical guardrails.
\end{abstract}



\maketitle

\section{Introduction}

Mental health challenges affect over 970 million people globally \cite{who2022mental}, with online communities like Reddit's r/depression\cite{reddit_depression} serving as critical spaces for peer support \& self-expression \cite{de2016discovering}. Computational approaches to understanding mental health in these spaces have predominantly relied on sentiment analysis \cite{de2013predicting}, topic modeling \cite{garg2023mental}, \& classification of diagnostic categories \cite{cohan2018smhd}. While valuable, these methods share a fundamental limitation: they treat mental health as a static classification problem rather than a \textit{dynamic trajectory} through psychological states.

Recent work by Alexander \& Wang \cite{alexander2023topological} demonstrated that Topological Data Analysis (TDA) can reveal meaningful structure in mental health discourse at a population level. However, their cross-sectional approach cannot capture how \textit{individual users} traverse psychological states over time, precisely what clinicians \& platform designers need to understand trajectory patterns \& design timely interventions \cite{tsakalidis2022identifying}.

\begin{figure*}[t]
    \centering
    \includegraphics[width=\textwidth]{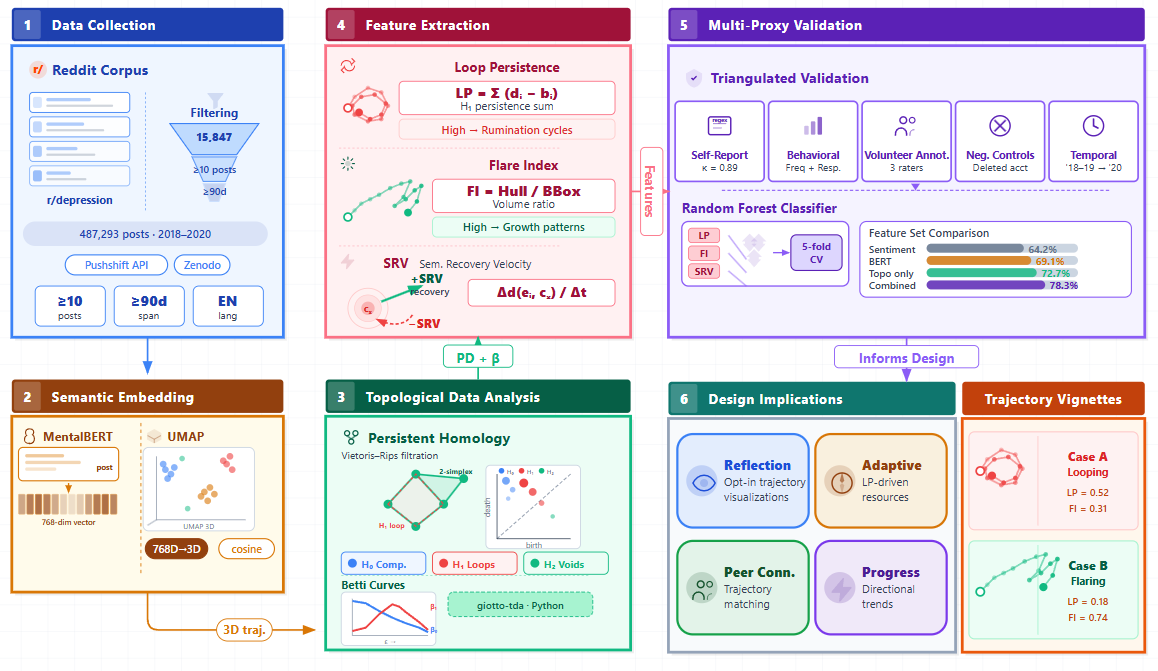}
    \caption{
    \textbf{ Overview of the Topology of Recovery pipeline. Reddit posts from r/depression (n = 15,847 users, 2018–2020) are encoded with MentalBERT and projected to 3D via UMAP. Vietoris–Rips filtration then extracts three topological features — Loop Persistence (LP), Flare Index (FI), and Semantic Recovery Velocity (SRV), which are validated against five proxies using a Random Forest classifier and translated into HCI design implications. Trajectory vignettes (right) illustrate two archetypal recovery patterns: looping (high LP) and flaring (high FI).}
    \Description{System architecture of The Topology of Recovery. The pipeline proceeds through six phases: (1) Data Collection — 15,847 longitudinal Reddit users (r/depression, ≥10 posts, ≥90-day span) are extracted via the Pushshift API; (2) Semantic Embedding — posts are encoded with MentalBERT (768-dim) and projected to 3D via UMAP; (3) Topological Data Analysis — Vietoris–Rips filtration computes persistent homology (H₀, H₁, H₂), yielding persistence diagrams and Betti curves via giotto-tda; (4) Feature Extraction — three interpretable topological features are derived: Loop Persistence (LP, rumination cycles), Flare Index (FI, growth expansion), and Semantic Recovery Velocity (SRV, directional momentum toward recovery); (5) Multi-Proxy Validation — features are validated against five proxies (self-report, behavioral signals, volunteer annotation, negative controls, and temporal holdout) using a Random Forest classifier (combined accuracy: 78.3\%); and (6) Design Implications — trajectory patterns inform four HCI design directions: reflective visualizations, adaptive resources, peer connection, and progress tracking. Representative trajectory vignettes illustrate the looping (high LP) and flaring (high FI) recovery archetypes.}
    }
    \label{fig:pipeline}
\end{figure*}

We propose a framework applying \textbf{persistent homology}, a mathematical tool from TDA to model individual users' longitudinal posting histories as \textit{trajectories through semantic embedding space}. In this context, \textit{semantic embedding space} refers to a high-dimensional vector space where each post is represented as a point based on its linguistic \& contextual meaning; posts with similar content appear closer together, enabling geometric analysis of how users' expressed concerns evolve over time. Our pipeline (Figure~\ref{fig:pipeline}) enables detection of topological features corresponding to meaningful patterns:

\begin{itemize}
    \item \textbf{Loops (H$_1$ features)}: When a user's semantic trajectory forms a topological loop, they are mathematically ``circling back'' to previously visited states—potentially indicating rumination or stagnation cycles \cite{nolen2008rethinking}. \textit{In plain terms: the user keeps returning to similar topics or emotional expressions.}
    \item \textbf{Flares}: Trajectories that extend into new regions of semantic space suggest exploration of new coping mechanisms or support-seeking behaviors. \textit{In plain terms: the user is discussing increasingly diverse topics over time.}
\end{itemize}

We introduce \textbf{Semantic Recovery Velocity (SRV)}, a metric quantifying the rate at which a user moves away from their computed "trauma center", the centroid of their early, distress-heavy posts. Unlike binary classifications, SRV provides a continuous measure of semantic movement. 

Our contributions are: 
\begin{enumerate}
    \item The first application of persistent homology to individual-level longitudinal mental health trajectories.
    \item A novel metric (SRV) \& topological interpretation framework. 
    \item Empirical validation with robustness analyses across embeddings \& parameters.
    \item Design implications with ethical guardrails for topology-aware platforms.
\end{enumerate}

\section{Related Work}

\subsection{Computational Mental Health Analysis}
The computational study of mental health in online communities has evolved from keyword-based detection \cite{de2013predicting} to deep learning approaches \cite{ji2022mentalbert}. Recent advances emphasize temporal dynamics: Tsakalidis et al. \cite{tsakalidis2022identifying} proposed identifying ``moments of change,'' while Morini et al. \cite{rossetti2024online} demonstrated non-linear posting patterns resembling clinical engagement models. Other work has focused on natural language processing for suicide risk screening \cite{coppersmith2018natural}, shared tasks for predicting suicide risk levels \cite{zirikly2019clpsych}, cognitive frameworks for detecting depression relapse \cite{agarwal2025redepress}, and prediction models for longitudinal trajectories of depression and anxiety \cite{fairweather2025prediction}. Research has also explored the perceived utility of digital self-tracking technologies for mental health \cite{kruzan2023perceived}. However, these approaches rely on discrete state classifications or scalar trajectories, missing the rich geometric structure of how users navigate psychological spaces, a gap our topological approach addresses.

\subsection{Topological Data Analysis in NLP and HCI}
TDA, particularly persistent homology, captures multi-scale data structure by identifying features (connected components, loops, voids) that persist across different scales of analysis \cite{chazal2021introduction}. Comprehensive roadmaps for computing persistent homology have made these methods more accessible to practitioners \cite{otter2017roadmap}. Recent surveys have unveiled the breadth of TDA applications in NLP \cite{uchendu2024unveiling}, while studies have examined topological properties of sentence embeddings \cite{das2021persistence}. In NLP, TDA has been applied to text classification \cite{zhu2013persistent}, topic modeling \cite{byrne2022topic}, and detecting LLM-generated text \cite{wei2025short}. Alexander \& Wang \cite{alexander2023topological} applied TDA to map mental health discourse at a population level, revealing relationships between psychiatric disorder communities. However, this cross-sectional, community-level analysis differs fundamentally from our individual-level, longitudinal approach. We track how \textit{single users} move through semantic space over months, not how communities relate to each other at one time point.

\subsection{Trajectory Analysis in Mental Health}
Clinical psychology recognizes recovery as non-linear \cite{de2019empirical}. Group-based trajectory modeling identifies subgroups following different patterns \cite{cukor2024longitudinal}, but assumes pre-defined trajectory shapes (linear, quadratic, etc.). Our topological approach is shape-agnostic, detecting meaningful structure without parametric assumptions about what trajectories ``should'' look like.

\section{Methods}

\subsection{Data Collection and Preprocessing}
We utilized the Reddit Mental Health Dataset \cite{low2020natural}, supplemented with longitudinal data via the Pushshift API \cite{baumgartner2020pushshift}. Inclusion criteria required users to have: (1) at least 10 posts spanning at least 90 days, (2) no deleted accounts, \& (3) posts primarily in English. Our corpus comprised 15,847 unique users with 487,293 posts spanning 2018–2020.

\subsection{Semantic Embedding and Trajectory Construction}
We embedded posts using MentalBERT \cite{ji2022mentalbert}, a BERT model fine-tuned on mental health corpora. For each user $u$ with chronologically ordered posts $\{p_1, \ldots, p_n\}$, we computed embeddings:
\[
\{e_1, \ldots, e_n\} \in \mathbb{R}^{768}
\]
Each embedding captures the semantic content, emotional tone, \& linguistic style of a post as a 768-dimensional vector.

We applied UMAP \cite{mcinnes2018umap} to project embeddings to $\mathbb{R}^{3}$, with parameters (n\_neighbors=15, min\_dist=0.1). This dimensionality reduction preserves local neighborhood structure while enabling persistent homology computation. To assess sensitivity, we conducted ablations varying these parameters \& using PCA as an alternative (see Section~\ref{sec:robustness}).

\subsection{Persistent Homology Computation}
\label{sec:tda_architecture}
Persistent homology is a method from algebraic topology that identifies topological features persisting across multiple scales. Our TDA pipeline operates as follows:

\textbf{Step 1: Vietoris-Rips Complex Construction.} For a user's 3D trajectory points, we construct simplicial complexes at increasing distance thresholds $\epsilon$. At each $\epsilon$, points within distance $\epsilon$ are connected, forming edges (1-simplices), triangles (2-simplices), \& higher-order structures.

\textbf{Step 2: Filtration.} As $\epsilon$ increases from 0 to $\infty$, topological features ``are born'' (appear) \& ``die'' (merge or fill in). We track three types of features:
\begin{itemize}
    \item \textbf{H$_0$ (Connected Components)}: Clusters of nearby points; many components at low $\epsilon$ merge into fewer as $\epsilon$ grows.
    \item \textbf{H$_1$ (Loops/Cycles)}: Circular structures in the trajectory; a loop is ``born'' when edges form a cycle \& ``dies'' when the interior is filled by triangles. \textit{These are our primary signal for ``looping back'' behavior.}
    \item \textbf{H$_2$ (Voids)}: 3D cavities enclosed by the trajectory; less common but can indicate complex exploration patterns.
\end{itemize}

\textbf{Step 3: Persistence Diagram.} The output is a diagram $D = \{(b_i, d_i, \dim_i)\}$ where each point represents a feature born at scale $b_i$, dying at $d_i$, with dimension $\dim_i$. Features far from the diagonal (high \textit{persistence} = $d_i - b_i$) represent robust topological structure rather than noise. We implemented this pipeline using \texttt{giotto-tda} \cite{tauzin2021giotto}.

\subsection{Topological Feature Extraction}
From persistence diagrams, we extract interpretable features:

\textbf{Loop Persistence (LP):} Total persistence of H$_1$ features:
\begin{equation}
\text{LP} = \sum_{(b,d,1) \in D} (d-b)
\end{equation}
\textit{Interpretation:} Higher LP indicates more prominent, persistent looping behavior—the user's trajectory contains robust cycles, suggesting repeated return to similar semantic regions.

\textbf{Flare Index (FI):} Ratio of trajectory convex hull volume to axis-aligned bounding box volume:
\begin{equation}
\text{FI} = \frac{\text{Volume}(\text{ConvexHull}(\tau))}{\text{Volume}(\text{BoundingBox}(\tau))}
\end{equation}
\textit{Interpretation:} Higher FI indicates the trajectory spreads out to fill more of its bounding space, suggesting broader exploration rather than constrained movement.

\subsection{Semantic Recovery Velocity (SRV)}
The \textbf{trauma center} $c_u$ is the centroid of a user's first $k=5$ posts' embeddings, representing their initial psychological ``location.'' SRV measures directional movement away from this center:
\begin{equation}
\text{SRV}_u = \frac{1}{n-k} \sum_{i=k+1}^{n} \frac{d(e'_i, c_u) - d(e'_{i-1}, c_u)}{\Delta t_i}
\end{equation}
where $d(\cdot, \cdot)$ is Euclidean distance \& $\Delta t_i$ is time between posts in days. Positive SRV indicates progressive movement away from initial distress-focused content; negative SRV indicates return toward it.

\subsection{Multi-Proxy Validation}
\label{sec:validation}
Given concerns about construct validity of self-reported labels \cite{harrigian2022then}, we employed triangulated validation with transparent collaboration protocols:

\textbf{(1) Pattern-matched self-reports:} Two researchers independently developed regex patterns for improvement phrases (e.g., ``I'm feeling better,'' ``things are improving,'' ``therapy is helping''). Patterns were refined through discussion until reaching consensus. We identified users where such phrases appeared in their final 20\% of posts but not their first 20\%, yielding 2,341 improved \& 3,892 non-improved users. Inter-rater reliability on a 200-post sample: $\kappa = 0.89$.

\textbf{(2) Behavioral proxies:} We computed: (a) posting frequency change (decrease in later periods, associated with reduced distress-driven posting in prior work); (b) community response rates (ratio of comments received). Agreement between pattern-matching \& behavioral proxies: Cohen's $\kappa = 0.41$ (moderate), indicating these capture related but distinct constructs.

\textbf{(3) Volunteer annotation:} We sampled 200 fully de identified, anonymized post sequences (all usernames, URLs, \& identifying details stripped). Three annotators per sequence rated whether trajectories showed "improvement," "stagnation," or "unclear" based on text progression alone. Annotators were recruited as pro bono volunteers from our research network, consisting of graduate students \& researchers with expertise in mental health informatics who volunteered their time for this analysis. Majority-vote labels achieved $kappa$ = 0.52 agreement with pattern-matched labels.

\textbf{(4) Negative controls:} We examined improvement patterns in users who later deleted accounts (potential false positives or transient users), finding only 12\% exhibited improvement patterns vs. 38\% in our retained improved group, supporting discriminant validity.

\textbf{(5) Temporal holdout:} We trained on 2018–2019 data \& tested on 2020 to assess temporal generalization.

\section{Results}

\subsection{Topological Signatures of Trajectory Patterns}
Users labeled as improved (via pattern-matched self-reports) showed significantly lower Loop Persistence compared to non-improved users (mean LP: 0.23 $\pm$ 0.14 vs. 0.41 $\pm$ 0.18, Cohen's $d = 0.58$, $p < 0.001$, Welch's $t$-test), indicating less semantic cycling. The Flare Index showed the opposite pattern: improved users exhibited higher FI (mean: 0.67 $\pm$ 0.19 vs. 0.52 $\pm$ 0.21, $d = 0.49$, $p < 0.001$), suggesting greater semantic exploration.

\subsection{SRV Validation}
SRV correlated with self-reported improvement ($r = 0.43$, $p < 0.001$, 95\% CI: [0.40, 0.46]). Users in the top SRV quartile were 3.2 times more likely to exhibit improvement markers (OR = 3.21, 95\% CI: 2.78–3.71). Crucially, SRV captured distinctions missed by sentiment analysis: among users with stable sentiment scores (within 0.1 SD across trajectory), SRV still differentiated improved from non-improved groups ($d = 0.34$, $p < 0.001$).

\subsection{Predictive Performance}
We trained Random Forest classifiers (100 trees, balanced class weights) using 5-fold stratified cross-validation. Table~\ref{tab:performance} reports mean $\pm$ standard deviation across folds. Temporal holdout (train 2018–19, test 2020) yielded 75.1\% accuracy (AUC = 0.79), demonstrating reasonable temporal generalization despite potential COVID-19 effects on 2020 data.

\begin{table*}[h]
\caption{Classification Performance for Self-Reported Improvement. All metrics reported as mean $\pm$ SD across 5 folds. Best results in bold.}
\label{tab:performance}
\begin{tabular}{lcccc}
\toprule
\textbf{Feature Set} & \textbf{Accuracy} & \textbf{F1} & \textbf{AUC} & \textbf{Precision} \\
\midrule
Sentiment (baseline) & 64.2 $\pm$ 2.1 & 0.61 $\pm$ 0.02 & 0.68 $\pm$ 0.02 & 0.63 $\pm$ 0.03 \\
BERT fine-tuned & 69.1 $\pm$ 1.8 & 0.66 $\pm$ 0.02 & 0.72 $\pm$ 0.02 & 0.68 $\pm$ 0.02 \\
Topic-shift metrics & 66.8 $\pm$ 2.3 & 0.64 $\pm$ 0.03 & 0.70 $\pm$ 0.02 & 0.65 $\pm$ 0.03 \\
Topological only & 72.7 $\pm$ 1.5 & 0.70 $\pm$ 0.02 & 0.76 $\pm$ 0.01 & 0.71 $\pm$ 0.02 \\
Combined & \textbf{78.3 $\pm$ 1.2} & \textbf{0.76 $\pm$ 0.01} & \textbf{0.82 $\pm$ 0.01} & \textbf{0.77 $\pm$ 0.02} \\
\bottomrule
\end{tabular}
\end{table*}

\subsection{Robustness Analyses}
\label{sec:robustness}
\textbf{UMAP sensitivity:} Varying n\_neighbors $\in \{10, 15, 20, 25, 30\}$ \& min\_dist $\in \{0.05, 0.1, 0.15, 0.2, 0.25\}$ across 5 random seeds (125 configurations), LP-improvement correlations remained stable ($r$ range: 0.38–0.47, mean: 0.42). PCA-based pipeline (retaining 95\% variance) yielded similar patterns ($r = 0.39$).

\textbf{Embedding ablation:} Results with BERT-base ($r = 0.36$) \& Sentence-Transformers/all-mpnet-base-v2 ($r = 0.40$) showed consistent, though attenuated, effects compared to MentalBERT ($r = 0.43$), confirming domain adaptation benefits but not dependency on a specific model.

\subsection{Illustrative Trajectory Vignettes}
To ground topological patterns in user experiences, we present anonymized cases (all identifying details removed):

\textbf{Case A (High LP, Low FI—``Looping''):} User posted 47 times over 14 months. Early posts focused on relationship difficulties; trajectory repeatedly returned to similar semantic regions despite discussing therapy attempts. Representative late post: \textit{``Same problems, different day. Nothing changes no matter what I try.''} LP = 0.52, FI = 0.31, SRV = -0.08.

\textbf{Case B (Low LP, High FI—"Flaring"):} User posted 38 times over 11 months. Initial crisis posts evolved toward therapy experiences, then medication discussions, then lifestyle changes, semantically distinct regions forming an outward trajectory. Representative late post: \textit{"Started running last month. Doesn't fix everything but I actually look forward to mornings now."} LP = 0.18, FI = 0.74, SRV = +0.31.

\textbf{Case C (Moderate LP, Moderate FI—``Mixed''):} User showed initial exploration (therapy posts) followed by return to crisis narratives after job loss, then gradual outward movement. This non-linear, context-dependent pattern captured by our temporal Betti curves illustrates why static classifications fail \& why trajectory based analysis is valuable.

\section{Discussion}

\subsection{Semantic Movement as Proxy, Not Diagnosis}
We emphasize that our topological features capture \textit{semantic movement patterns}, not clinical recovery. High SRV indicates a user's posts are exploring different semantic territory over time; this \textit{correlates with} but does not \textit{constitute} psychological improvement. Users may ``flare'' into new regions for many reasons (new stressors, changed life circumstances, platform migration) unrelated to wellbeing. Our claims are observational as topological patterns are associated with self-reported \& crowd-validated improvement markers, warranting further investigation with clinical validation.

\subsection{Design Implications: Opportunities \& Risks}
\label{sec:design}
Our findings suggest design directions, but each requires careful consideration of user reception \& potential harms. Previous work has established important ethical frameworks for inferring mental health states from social media \cite{chancellor2019taxonomy}:

\textbf{Topology-Aware Reflection Tools.} Opt-in visualizations could show users their semantic trajectory, prompting reflection when patterns suggest stagnation.
\begin{itemize}
    \item \textit{When helpful:} Users actively seeking self-insight, those in therapeutic relationships who could discuss patterns with clinicians, or researchers/moderators analyzing community health.
    \item \textit{Risk of harm:} Users may feel \textbf{monitored or surveilled}, especially if features are opt-out rather than opt-in. Trajectory visualizations could trigger anxiety (``Am I stuck?'') or become obsessive self-monitoring tools. Users in acute crisis may find such feedback destabilizing.
    \item \textit{Mitigation:} Explicit opt-in consent with clear framing that patterns are exploratory, not diagnostic. Include prominent disclaimers \& easy opt-out. Consider gating behind clinical/peer support contexts rather than direct-to-user deployment.
\end{itemize}

\textbf{Resource Suggestion.} High Loop Persistence could trigger suggestions for diverse resources, e.g., ``Users who explored similar topics found these posts helpful.''
\begin{itemize}
    \item \textit{When helpful:} Users expressing openness to new approaches, those whose posting patterns suggest they may benefit from diverse perspectives.
    \item \textit{Risk of harm:} Users may feel \textbf{labeled or categorized} (``The algorithm thinks I'm stuck''), leading to stigmatization or reduced sense of agency. Suggestions may feel patronizing or intrusive.
    \item \textit{Mitigation:} Frame suggestions as community features (``others found helpful'') rather than individual inference. Never explicitly reveal the user's topological classification. Allow users to indicate they don't want suggestions.
\end{itemize}

\textbf{Peer Connection.} Users with complementary profiles could be suggested as peer connections.
\begin{itemize}
    \item \textit{When helpful:} Connecting someone in a ``looping'' pattern with someone who navigated similar challenges could provide hope \& practical guidance.
    \item \textit{Risk of harm:} Privacy concerns about why the match was made. Power imbalances if one user feels positioned as ``recovered'' \& another as ``struggling.'' Potential for unhealthy dynamics if matching is perceived as clinical rather than social.
    \item \textit{Mitigation:} Human moderator oversight for all suggested connections. Never reveal matching rationale. Frame as interest-based rather than trajectory-based matching.
\end{itemize}

\textbf{Progress Indicators.} SRV could power continuous indicators showing semantic movement.
\begin{itemize}
    \item \textit{When helpful:} Users who find quantified feedback motivating, those in formal recovery programs where progress tracking is part of treatment.
    \item \textit{Risk of harm:} \textbf{Gamification of recovery} could lead to posting behavior optimized for metrics rather than authentic expression. Negative SRV could be demoralizing. Users may feel judged by their ``score.''
    \item \textit{Mitigation:} If implemented, show only directional trends (``exploring new topics'') rather than numeric scores. Never frame as improvement/decline. Consider showing this detail only to clinicians/supporters rather than users directly.
\end{itemize}

\textbf{Overarching Principle:} Any topology-based feature should be designed with the assumption that some users will feel uncomfortable being analyzed. The appropriate deployment context is likely \textit{not} direct-to-user consumer features, but rather: (1) research tools for understanding community dynamics, (2) opt-in features within clinical or peer support programs with human oversight, or (3) moderator dashboards for identifying users who might benefit from outreach (with human review before any action).

\subsection{Ethical Considerations and Deployment Guardrails}
Analyzing mental health data carries profound responsibilities. We used only publicly posted, anonymized data \& make no individual-level clinical claims.

\textbf{Critical deployment constraints:} Any implementation of topology-based tools \textbf{must not} be deployed without: (1) clinical oversight \& validation studies with IRB approval; (2) participatory design with affected communities, including people with lived experience of depression, \& platform moderators; (3) explicit opt-in consent with clear explanations of limitations; (4) human-in-the-loop review for any intervention triggers; (5) robust privacy protections exceeding current standards; \& (6) mechanisms for users to contest, correct, or opt-out of any inferences.

We strongly caution against using these methods for risk prediction, content moderation decisions, or any consequential individual-level determinations without extensive clinical validation. Topological patterns should inform research \& participatory design exploration, not automated interventions.

\subsection{Limitations}
Our proxy labels, while triangulated across five validation approaches, remain imperfect; clinical validation with ground-truth outcomes is essential future work. UMAP introduces stochasticity that, while shown robust in aggregate statistics, affects individual trajectory shapes, we recommend ensemble approaches for any individual-level applications. Cross-community generalization (to other subreddits, platforms, or conditions) requires dedicated study. The 2018–2020 time frame may not reflect current community dynamics, particularly post-COVID changes in online mental health discourse.

\section{Conclusion}
We introduced a framework applying persistent homology to map individual mental health trajectories in online communities. By conceptualizing posting patterns as trajectories through semantic space, we revealed topological signatures - loops \& flares associated with self-reported improvement patterns across multiple validation approaches. Our Semantic Recovery Velocity metric provides a continuous measure of semantic movement, with robustness demonstrated across embeddings \& parameters. With appropriate caution about the observational nature of our claims, critical ethical guardrails, \& careful consideration of user reception, this work opens methodological directions for HCI researchers studying mental health technologies.

\begin{acks}
We thank the volunteer annotators from our research network for their pro bono contributions to the validation process. We would also like to acknowledge that the teaser illustration (Figure \ref{fig:teaser}) was generated using Gemini for reference \& conceptual visualization purposes.
\end{acks}

\bibliographystyle{ACM-Reference-Format}
\bibliography{references}


\begin{thebibliography}{31}


\ifx \showCODEN    \undefined \def \showCODEN     #1{\unskip}     \fi
\ifx \showISBNx    \undefined \def \showISBNx     #1{\unskip}     \fi
\ifx \showISBNxiii \undefined \def \showISBNxiii  #1{\unskip}     \fi
\ifx \showISSN     \undefined \def \showISSN      #1{\unskip}     \fi
\ifx \showLCCN     \undefined \def \showLCCN      #1{\unskip}     \fi
\ifx \shownote     \undefined \def \shownote      #1{#1}          \fi
\ifx \showarticletitle \undefined \def \showarticletitle #1{#1}   \fi
\ifx \showURL      \undefined \def \showURL       {\relax}        \fi
\providecommand\bibfield[2]{#2}
\providecommand\bibinfo[2]{#2}
\providecommand\natexlab[1]{#1}
\providecommand\showeprint[2][]{arXiv:#2}

\bibitem[Agarwal et~al\mbox{.}(2025)]%
        {agarwal2025redepress}
\bibfield{author}{\bibinfo{person}{Aakash~Kumar Agarwal}, \bibinfo{person}{Saprativa Bhattacharjee}, \bibinfo{person}{Mauli Rastogi}, \bibinfo{person}{Jemima~S. Jacob}, \bibinfo{person}{Biplab Banerjee}, \bibinfo{person}{Rashmi Gupta}, {and} \bibinfo{person}{Pushpak Bhattacharyya}.} \bibinfo{year}{2025}\natexlab{}.
\newblock \bibinfo{title}{ReDepress: A Cognitive Framework for Detecting Depression Relapse from Social Media}.
\newblock
\showeprint[arxiv]{2509.17991}~[cs.CL]
\urldef\tempurl%
\url{https://arxiv.org/abs/2509.17991}
\showURL{%
\tempurl}


\bibitem[Alexander and Wang(2023)]%
        {alexander2023topological}
\bibfield{author}{\bibinfo{person}{Andrew Alexander} {and} \bibinfo{person}{Hongbin Wang}.} \bibinfo{year}{2023}\natexlab{}.
\newblock \bibinfo{title}{Topological Data Mapping of Online Hate Speech, Misinformation, and General Mental Health: A Large Language Model Based Study}.
\newblock
\showeprint[arxiv]{2309.13098}~[cs.LG]
\urldef\tempurl%
\url{https://arxiv.org/abs/2309.13098}
\showURL{%
\tempurl}


\bibitem[Baumgartner et~al\mbox{.}(2020)]%
        {baumgartner2020pushshift}
\bibfield{author}{\bibinfo{person}{Jason Baumgartner}, \bibinfo{person}{Savvas Zannettou}, \bibinfo{person}{Brian Keegan}, \bibinfo{person}{Megan Squire}, {and} \bibinfo{person}{Jeremy Blackburn}.} \bibinfo{year}{2020}\natexlab{}.
\newblock \bibinfo{title}{The Pushshift Reddit Dataset}.
\newblock
\showeprint[arxiv]{2001.08435}~[cs.SI]
\urldef\tempurl%
\url{https://arxiv.org/abs/2001.08435}
\showURL{%
\tempurl}


\bibitem[Byrne et~al\mbox{.}(2022)]%
        {byrne2022topic}
\bibfield{author}{\bibinfo{person}{Ciar{\'a}n Byrne}, \bibinfo{person}{Danijela Horak}, \bibinfo{person}{Karo Moilanen}, {and} \bibinfo{person}{Amandla Mabona}.} \bibinfo{year}{2022}\natexlab{}.
\newblock \showarticletitle{Topic Modeling With Topological Data Analysis}. In \bibinfo{booktitle}{\emph{Proceedings of the 2022 Conference on Empirical Methods in Natural Language Processing}}, \bibfield{editor}{\bibinfo{person}{Yoav Goldberg}, \bibinfo{person}{Zornitsa Kozareva}, {and} \bibinfo{person}{Yue Zhang}} (Eds.). \bibinfo{publisher}{Association for Computational Linguistics}, \bibinfo{address}{Abu Dhabi, United Arab Emirates}, \bibinfo{pages}{11514--11533}.
\newblock
\href{https://doi.org/10.18653/v1/2022.emnlp-main.792}{doi:\nolinkurl{10.18653/v1/2022.emnlp-main.792}}


\bibitem[Chancellor et~al\mbox{.}(2019)]%
        {chancellor2019taxonomy}
\bibfield{author}{\bibinfo{person}{Stevie Chancellor}, \bibinfo{person}{Michael~L. Birnbaum}, \bibinfo{person}{Eric~D. Caine}, \bibinfo{person}{Vincent M.~B. Silenzio}, {and} \bibinfo{person}{Munmun De~Choudhury}.} \bibinfo{year}{2019}\natexlab{}.
\newblock \showarticletitle{A Taxonomy of Ethical Tensions in Inferring Mental Health States from Social Media}. In \bibinfo{booktitle}{\emph{Proceedings of the Conference on Fairness, Accountability, and Transparency}} (Atlanta, GA, USA) \emph{(\bibinfo{series}{FAT* '19})}. \bibinfo{publisher}{Association for Computing Machinery}, \bibinfo{address}{New York, NY, USA}, \bibinfo{pages}{79–88}.
\newblock
\showISBNx{9781450361255}
\href{https://doi.org/10.1145/3287560.3287587}{doi:\nolinkurl{10.1145/3287560.3287587}}


\bibitem[Chazal and Michel(2021)]%
        {chazal2021introduction}
\bibfield{author}{\bibinfo{person}{Frédéric Chazal} {and} \bibinfo{person}{Bertrand Michel}.} \bibinfo{year}{2021}\natexlab{}.
\newblock \bibinfo{title}{An introduction to Topological Data Analysis: fundamental and practical aspects for data scientists}.
\newblock
\showeprint[arxiv]{1710.04019}~[math.ST]
\urldef\tempurl%
\url{https://arxiv.org/abs/1710.04019}
\showURL{%
\tempurl}


\bibitem[Cohan et~al\mbox{.}(2018)]%
        {cohan2018smhd}
\bibfield{author}{\bibinfo{person}{Arman Cohan}, \bibinfo{person}{Bart Desmet}, \bibinfo{person}{Andrew Yates}, \bibinfo{person}{Luca Soldaini}, \bibinfo{person}{Sean MacAvaney}, {and} \bibinfo{person}{Nazli Goharian}.} \bibinfo{year}{2018}\natexlab{}.
\newblock \bibinfo{title}{SMHD: A Large-Scale Resource for Exploring Online Language Usage for Multiple Mental Health Conditions}.
\newblock
\showeprint[arxiv]{1806.05258}~[cs.CL]
\urldef\tempurl%
\url{https://arxiv.org/abs/1806.05258}
\showURL{%
\tempurl}


\bibitem[Coppersmith et~al\mbox{.}(2018)]%
        {coppersmith2018natural}
\bibfield{author}{\bibinfo{person}{Glen Coppersmith}, \bibinfo{person}{Ryan Leary}, \bibinfo{person}{Patrick Crutchley}, {and} \bibinfo{person}{Alex Fine}.} \bibinfo{year}{2018}\natexlab{}.
\newblock \showarticletitle{Natural Language Processing of Social Media as Screening for Suicide Risk}.
\newblock \bibinfo{journal}{\emph{Biomedical Informatics Insights}}  \bibinfo{volume}{10} (\bibinfo{year}{2018}), \bibinfo{pages}{1178222618792860}.
\newblock
\showeprint{https://doi.org/10.1177/1178222618792860}
\href{https://doi.org/10.1177/1178222618792860}{doi:\nolinkurl{10.1177/1178222618792860}}
\newblock
\shownote{PMID: 30158822}.


\bibitem[Cukor et~al\mbox{.}(2024)]%
        {cukor2024longitudinal}
\bibfield{author}{\bibinfo{person}{J. Cukor}, \bibinfo{person}{Z. Xu}, \bibinfo{person}{V. Vekaria}, {et~al\mbox{.}}} \bibinfo{year}{2024}\natexlab{}.
\newblock \showarticletitle{Longitudinal trajectories of symptom change during antidepressant treatment among managed care patients with depression and anxiety}.
\newblock \bibinfo{journal}{\emph{npj Mental Health Research}} \bibinfo{volume}{3}, \bibinfo{number}{58} (\bibinfo{year}{2024}).
\newblock
\href{https://doi.org/10.1038/s44184-024-00104-8}{doi:\nolinkurl{10.1038/s44184-024-00104-8}}


\bibitem[Das et~al\mbox{.}(2021)]%
        {das2021persistence}
\bibfield{author}{\bibinfo{person}{Shouman Das}, \bibinfo{person}{Syed~A. Haque}, {and} \bibinfo{person}{Md.~Iftekhar Tanveer}.} \bibinfo{year}{2021}\natexlab{}.
\newblock \bibinfo{title}{Persistence Homology of TEDtalk: Do Sentence Embeddings Have a Topological Shape?}
\newblock
\showeprint[arxiv]{2103.14131}~[cs.LG]
\urldef\tempurl%
\url{https://arxiv.org/abs/2103.14131}
\showURL{%
\tempurl}


\bibitem[De~Choudhury et~al\mbox{.}(2021)]%
        {de2013predicting}
\bibfield{author}{\bibinfo{person}{Munmun De~Choudhury}, \bibinfo{person}{Michael Gamon}, \bibinfo{person}{Scott Counts}, {and} \bibinfo{person}{Eric Horvitz}.} \bibinfo{year}{2021}\natexlab{}.
\newblock \showarticletitle{Predicting Depression via Social Media}.
\newblock \bibinfo{journal}{\emph{Proceedings of the International AAAI Conference on Web and Social Media}} \bibinfo{volume}{7}, \bibinfo{number}{1} (\bibinfo{date}{Aug.} \bibinfo{year}{2021}), \bibinfo{pages}{128--137}.
\newblock
\href{https://doi.org/10.1609/icwsm.v7i1.14432}{doi:\nolinkurl{10.1609/icwsm.v7i1.14432}}


\bibitem[De~Choudhury et~al\mbox{.}(2016)]%
        {de2016discovering}
\bibfield{author}{\bibinfo{person}{Munmun De~Choudhury}, \bibinfo{person}{Emre Kiciman}, \bibinfo{person}{Mark Dredze}, \bibinfo{person}{Glen Coppersmith}, {and} \bibinfo{person}{Mrinal Kumar}.} \bibinfo{year}{2016}\natexlab{}.
\newblock \showarticletitle{Discovering Shifts to Suicidal Ideation from Mental Health Content in Social Media}. In \bibinfo{booktitle}{\emph{Proceedings of the 2016 CHI Conference on Human Factors in Computing Systems}} (San Jose, California, USA) \emph{(\bibinfo{series}{CHI '16})}. \bibinfo{publisher}{Association for Computing Machinery}, \bibinfo{address}{New York, NY, USA}, \bibinfo{pages}{2098–2110}.
\newblock
\showISBNx{9781450333627}
\href{https://doi.org/10.1145/2858036.2858207}{doi:\nolinkurl{10.1145/2858036.2858207}}


\bibitem[de~Zwart et~al\mbox{.}(2019)]%
        {de2019empirical}
\bibfield{author}{\bibinfo{person}{P.~L. de Zwart}, \bibinfo{person}{B.~F. Jeronimus}, {and} \bibinfo{person}{P. de Jonge}.} \bibinfo{year}{2019}\natexlab{}.
\newblock \showarticletitle{Empirical evidence for definitions of episode, remission, recovery, relapse and recurrence in depression: a systematic review}.
\newblock \bibinfo{journal}{\emph{Epidemiology and Psychiatric Sciences}} \bibinfo{volume}{28}, \bibinfo{number}{5} (\bibinfo{year}{2019}), \bibinfo{pages}{544–562}.
\newblock
\href{https://doi.org/10.1017/S2045796018000227}{doi:\nolinkurl{10.1017/S2045796018000227}}


\bibitem[Fairweather et~al\mbox{.}(2026)]%
        {fairweather2025prediction}
\bibfield{author}{\bibinfo{person}{Sophie~J. Fairweather}, \bibinfo{person}{Holly Fraser}, \bibinfo{person}{Natalie Lam}, \bibinfo{person}{Simon Gilbody}, \bibinfo{person}{Lewis~W. Paton}, \bibinfo{person}{Hannah~J. Jones}, {and} \bibinfo{person}{Golam~M. Khandaker}.} \bibinfo{year}{2026}\natexlab{}.
\newblock \showarticletitle{Prediction models for longitudinal trajectories of depression and anxiety: a systematic review}.
\newblock \bibinfo{journal}{\emph{Journal of Affective Disorders}}  \bibinfo{volume}{401} (\bibinfo{year}{2026}), \bibinfo{pages}{121255}.
\newblock
\showISSN{0165-0327}
\href{https://doi.org/10.1016/j.jad.2026.121255}{doi:\nolinkurl{10.1016/j.jad.2026.121255}}


\bibitem[Garg(2023)]%
        {garg2023mental}
\bibfield{author}{\bibinfo{person}{Muskan Garg}.} \bibinfo{year}{2023}\natexlab{}.
\newblock \showarticletitle{Mental Health Analysis in Social Media Posts: A Survey}.
\newblock \bibinfo{journal}{\emph{Archives of Computational Methods in Engineering}} \bibinfo{volume}{30}, \bibinfo{number}{3} (\bibinfo{year}{2023}), \bibinfo{pages}{1819--1842}.
\newblock
\href{https://doi.org/10.1007/s11831-022-09863-z}{doi:\nolinkurl{10.1007/s11831-022-09863-z}}


\bibitem[Harrigian and Dredze(2022)]%
        {harrigian2022then}
\bibfield{author}{\bibinfo{person}{Keith Harrigian} {and} \bibinfo{person}{Mark Dredze}.} \bibinfo{year}{2022}\natexlab{}.
\newblock \bibinfo{title}{Then and Now: Quantifying the Longitudinal Validity of Self-Disclosed Depression Diagnoses}.
\newblock
\showeprint[arxiv]{2206.11155}~[cs.LG]
\urldef\tempurl%
\url{https://arxiv.org/abs/2206.11155}
\showURL{%
\tempurl}


\bibitem[Ji et~al\mbox{.}(2022)]%
        {ji2022mentalbert}
\bibfield{author}{\bibinfo{person}{Shaoxiong Ji}, \bibinfo{person}{Tianlin Zhang}, \bibinfo{person}{Luna Ansari}, \bibinfo{person}{Jie Fu}, \bibinfo{person}{Prayag Tiwari}, {and} \bibinfo{person}{Erik Cambria}.} \bibinfo{year}{2022}\natexlab{}.
\newblock \showarticletitle{{M}ental{BERT}: Publicly Available Pretrained Language Models for Mental Healthcare}.
\newblock  (\bibinfo{date}{June} \bibinfo{year}{2022}), \bibinfo{pages}{7184--7190}.
\newblock
\urldef\tempurl%
\url{https://aclanthology.org/2022.lrec-1.778/}
\showURL{%
\tempurl}


\bibitem[Kruzan et~al\mbox{.}(2023)]%
        {kruzan2023perceived}
\bibfield{author}{\bibinfo{person}{Kaylee~Payne Kruzan}, \bibinfo{person}{Ada Ng}, \bibinfo{person}{Colleen Stiles-Shields}, \bibinfo{person}{Emily~G Lattie}, \bibinfo{person}{David~C. Mohr}, {and} \bibinfo{person}{Madhu Reddy}.} \bibinfo{year}{2023}\natexlab{}.
\newblock \showarticletitle{The Perceived Utility of Smartphone and Wearable Sensor Data in Digital Self-tracking Technologies for Mental Health}. In \bibinfo{booktitle}{\emph{Proceedings of the 2023 CHI Conference on Human Factors in Computing Systems}} (Hamburg, Germany) \emph{(\bibinfo{series}{CHI '23})}. \bibinfo{publisher}{Association for Computing Machinery}, \bibinfo{address}{New York, NY, USA}, Article \bibinfo{articleno}{88}, \bibinfo{numpages}{16}~pages.
\newblock
\showISBNx{9781450394215}
\href{https://doi.org/10.1145/3544548.3581209}{doi:\nolinkurl{10.1145/3544548.3581209}}


\bibitem[Low et~al\mbox{.}(2020)]%
        {low2020natural}
\bibfield{author}{\bibinfo{person}{Daniel~M Low}, \bibinfo{person}{Laurie Rumker}, \bibinfo{person}{Tanya Talkar}, \bibinfo{person}{John Torous}, \bibinfo{person}{Guillermo Cecchi}, {and} \bibinfo{person}{Satrajit~S Ghosh}.} \bibinfo{year}{2020}\natexlab{}.
\newblock \showarticletitle{Natural Language Processing Reveals Vulnerable Mental Health Support Groups and Heightened Health Anxiety on Reddit During COVID-19: Observational Study}.
\newblock \bibinfo{journal}{\emph{J Med Internet Res}} \bibinfo{volume}{22}, \bibinfo{number}{10} (\bibinfo{date}{12 Oct} \bibinfo{year}{2020}), \bibinfo{pages}{e22635}.
\newblock
\showISSN{1438-8871}
\href{https://doi.org/10.2196/22635}{doi:\nolinkurl{10.2196/22635}}


\bibitem[McInnes et~al\mbox{.}(2020)]%
        {mcinnes2018umap}
\bibfield{author}{\bibinfo{person}{Leland McInnes}, \bibinfo{person}{John Healy}, {and} \bibinfo{person}{James Melville}.} \bibinfo{year}{2020}\natexlab{}.
\newblock \bibinfo{title}{UMAP: Uniform Manifold Approximation and Projection for Dimension Reduction}.
\newblock
\showeprint[arxiv]{1802.03426}~[stat.ML]
\urldef\tempurl%
\url{https://arxiv.org/abs/1802.03426}
\showURL{%
\tempurl}


\bibitem[Morini et~al\mbox{.}(2025)]%
        {rossetti2024online}
\bibfield{author}{\bibinfo{person}{Virginia Morini}, \bibinfo{person}{Salvatore Citraro}, \bibinfo{person}{Elena Sajno}, \bibinfo{person}{Maria Sansoni}, \bibinfo{person}{Giuseppe Riva}, \bibinfo{person}{Massimo Stella}, {and} \bibinfo{person}{Giulio Rossetti}.} \bibinfo{year}{2025}\natexlab{}.
\newblock \bibinfo{title}{Online posting effects: Unveiling the non-linear journeys of users in depression communities on Reddit}.
\newblock
\showeprint[arxiv]{2311.17684}~[cs.SI]
\urldef\tempurl%
\url{https://arxiv.org/abs/2311.17684}
\showURL{%
\tempurl}


\bibitem[Nolen-Hoeksema et~al\mbox{.}(2008)]%
        {nolen2008rethinking}
\bibfield{author}{\bibinfo{person}{Susan Nolen-Hoeksema}, \bibinfo{person}{Blair~E. Wisco}, {and} \bibinfo{person}{Sonja Lyubomirsky}.} \bibinfo{year}{2008}\natexlab{}.
\newblock \showarticletitle{Rethinking Rumination}.
\newblock \bibinfo{journal}{\emph{Perspectives on Psychological Science}} \bibinfo{volume}{3}, \bibinfo{number}{5} (\bibinfo{year}{2008}), \bibinfo{pages}{400--424}.
\newblock
\showeprint{https://doi.org/10.1111/j.1745-6924.2008.00088.x}
\href{https://doi.org/10.1111/j.1745-6924.2008.00088.x}{doi:\nolinkurl{10.1111/j.1745-6924.2008.00088.x}}
\newblock
\shownote{PMID: 26158958}.


\bibitem[Otter et~al\mbox{.}(2017)]%
        {otter2017roadmap}
\bibfield{author}{\bibinfo{person}{Nina Otter}, \bibinfo{person}{Mason~A Porter}, \bibinfo{person}{Ulrike Tillmann}, \bibinfo{person}{Peter Grindrod}, {and} \bibinfo{person}{Heather~A Harrington}.} \bibinfo{year}{2017}\natexlab{}.
\newblock \showarticletitle{A roadmap for the computation of persistent homology}.
\newblock \bibinfo{journal}{\emph{EPJ Data Science}} \bibinfo{volume}{6}, \bibinfo{number}{1} (\bibinfo{date}{Aug.} \bibinfo{year}{2017}).
\newblock
\showISSN{2193-1127}
\href{https://doi.org/10.1140/epjds/s13688-017-0109-5}{doi:\nolinkurl{10.1140/epjds/s13688-017-0109-5}}


\bibitem[{r/depression Community}(2026)]%
        {reddit_depression}
\bibfield{author}{\bibinfo{person}{{r/depression Community}}.} \bibinfo{year}{2026}\natexlab{}.
\newblock \bibinfo{title}{Depression: A place to vent, to support, and to share information}.
\newblock \bibinfo{howpublished}{\url{https://www.reddit.com/r/depression/}}.
\newblock
\newblock
\shownote{Accessed: February 27, 2026}.


\bibitem[Tauzin et~al\mbox{.}(2021)]%
        {tauzin2021giotto}
\bibfield{author}{\bibinfo{person}{Guillaume Tauzin}, \bibinfo{person}{Umberto Lupo}, \bibinfo{person}{Lewis Tunstall}, \bibinfo{person}{Julian~Burella Pérez}, \bibinfo{person}{Matteo Caorsi}, \bibinfo{person}{Wojciech Reise}, \bibinfo{person}{Anibal Medina-Mardones}, \bibinfo{person}{Alberto Dassatti}, {and} \bibinfo{person}{Kathryn Hess}.} \bibinfo{year}{2021}\natexlab{}.
\newblock \bibinfo{title}{giotto-tda: A Topological Data Analysis Toolkit for Machine Learning and Data Exploration}.
\newblock
\showeprint[arxiv]{2004.02551}~[cs.LG]
\urldef\tempurl%
\url{https://arxiv.org/abs/2004.02551}
\showURL{%
\tempurl}


\bibitem[Tsakalidis et~al\mbox{.}(2022)]%
        {tsakalidis2022identifying}
\bibfield{author}{\bibinfo{person}{Adam Tsakalidis}, \bibinfo{person}{Federico Nanni}, \bibinfo{person}{Anthony Hills}, \bibinfo{person}{Jenny Chim}, \bibinfo{person}{Jiayu Song}, {and} \bibinfo{person}{Maria Liakata}.} \bibinfo{year}{2022}\natexlab{}.
\newblock \bibinfo{title}{Identifying Moments of Change from Longitudinal User Text}.
\newblock
\showeprint[arxiv]{2205.05593}~[cs.CL]
\urldef\tempurl%
\url{https://arxiv.org/abs/2205.05593}
\showURL{%
\tempurl}


\bibitem[Uchendu and Le(2025)]%
        {uchendu2024unveiling}
\bibfield{author}{\bibinfo{person}{Adaku Uchendu} {and} \bibinfo{person}{Thai Le}.} \bibinfo{year}{2025}\natexlab{}.
\newblock \bibinfo{title}{Unveiling Topological Structures from Language: A Survey of Topological Data Analysis Applications in NLP}.
\newblock
\showeprint[arxiv]{2411.10298}~[cs.CL]
\urldef\tempurl%
\url{https://arxiv.org/abs/2411.10298}
\showURL{%
\tempurl}


\bibitem[Wei et~al\mbox{.}(2025)]%
        {wei2025short}
\bibfield{author}{\bibinfo{person}{Dongjun Wei}, \bibinfo{person}{Minjia Mao}, \bibinfo{person}{Xiao Fang}, {and} \bibinfo{person}{Michael Chau}.} \bibinfo{year}{2025}\natexlab{}.
\newblock \bibinfo{title}{Short-PHD: Detecting Short LLM-generated Text with Topological Data Analysis After Off-topic Content Insertion}.
\newblock
\showeprint[arxiv]{2504.02873}~[cs.CL]
\urldef\tempurl%
\url{https://arxiv.org/abs/2504.02873}
\showURL{%
\tempurl}


\bibitem[{World Health Organization}(2022)]%
        {who2022mental}
\bibfield{author}{\bibinfo{person}{{World Health Organization}}.} \bibinfo{year}{2022}\natexlab{}.
\newblock \bibinfo{title}{Mental disorders}.
\newblock \bibinfo{howpublished}{\url{https://www.who.int/news-room/fact-sheets/detail/mental-disorders}}.
\newblock


\bibitem[Zhu(2013)]%
        {zhu2013persistent}
\bibfield{author}{\bibinfo{person}{Xiaojin Zhu}.} \bibinfo{year}{2013}\natexlab{}.
\newblock \showarticletitle{Persistent homology: an introduction and a new text representation for natural language processing}. In \bibinfo{booktitle}{\emph{Proceedings of the Twenty-Third International Joint Conference on Artificial Intelligence}} (Beijing, China) \emph{(\bibinfo{series}{IJCAI '13})}. \bibinfo{publisher}{AAAI Press}, \bibinfo{pages}{1953–1959}.
\newblock
\showISBNx{9781577356332}


\bibitem[Zirikly et~al\mbox{.}(2019)]%
        {zirikly2019clpsych}
\bibfield{author}{\bibinfo{person}{Ayah Zirikly}, \bibinfo{person}{Philip Resnik}, \bibinfo{person}{{\"O}zlem Uzuner}, {and} \bibinfo{person}{Kristy Hollingshead}.} \bibinfo{year}{2019}\natexlab{}.
\newblock \showarticletitle{{CLP}sych 2019 Shared Task: Predicting the Degree of Suicide Risk in {R}eddit Posts}. In \bibinfo{booktitle}{\emph{Proceedings of the Sixth Workshop on Computational Linguistics and Clinical Psychology}}, \bibfield{editor}{\bibinfo{person}{Kate Niederhoffer}, \bibinfo{person}{Kristy Hollingshead}, \bibinfo{person}{Philip Resnik}, \bibinfo{person}{Rebecca Resnik}, {and} \bibinfo{person}{Kate Loveys}} (Eds.). \bibinfo{publisher}{Association for Computational Linguistics}, \bibinfo{address}{Minneapolis, Minnesota}, \bibinfo{pages}{24--33}.
\newblock
\href{https://doi.org/10.18653/v1/W19-3003}{doi:\nolinkurl{10.18653/v1/W19-3003}}


\end{thebibliography}

\end{document}